\documentclass[a4paper,12pt]{article}
\usepackage{amsfonts,amsthm,amsmath,amssymb,graphicx,hyperref,youngtab,color,tcolorbox}
\numberwithin{equation}{section}
\usepackage{bbold}

\begin{document}
\newtheorem{definition}{Definition}[section]
\newcommand{\be}{\begin{equation}}
\newcommand{\ee}{\end{equation}}
\newcommand{\bea}{\begin{eqnarray}}
\newcommand{\eea}{\end{eqnarray}}
\newcommand{\LE}{\left[}
\newcommand{\me}{\mathrm{e}}
\newcommand{\R}{\right]}
\newcommand{\nn}{\nonumber}
\newcommand{\Tr}{\text{Tr}}
\newcommand{\N}{\mathcal{N}}
\newcommand{\G}{\Gamma}
\newcommand{\vf}{\varphi}
\newcommand{\LL}{\mathcal{L}}
\newcommand{\Op}{\mathcal{O}}
\newcommand{\HH}{\mathcal{H}}
\newcommand{\arctanh}{\text{arctanh}}
\newcommand{\up}{\uparrow}
\newcommand{\down}{\downarrow}
\newcommand{\ket}[1]{\left| #1 \right>}
\newcommand{\bra}[1]{\left< #1 \right|}
\newcommand{\ketbra}[1]{\left|#1\right>\left<#1\right|}
\newcommand{\rd}{\partial}
\newcommand{\de}{\partial}
\newcommand{\ba}{\begin{eqnarray}}
\newcommand{\ea}{\end{eqnarray}}
\newcommand{\db}{\bar{\partial}}
\newcommand{\we}{\wedge}
\newcommand{\ca}{\mathcal}
\newcommand{\lr}{\leftrightarrow}
\newcommand{\f}{\frac}
\newcommand{\s}{\sqrt}
\newcommand{\vp}{\varphi}
\newcommand{\hvp}{\hat{\varphi}}
\newcommand{\tvp}{\tilde{\varphi}}
\newcommand{\tp}{\tilde{\phi}}
\newcommand{\ti}{\tilde}
\newcommand{\ap}{\alpha}
\newcommand{\pr}{\propto}
\newcommand{\mb}{\mathbf}
\newcommand{\ddd}{\cdot\cdot\cdot}
\newcommand{\no}{\nonumber \\}
\newcommand{\la}{\langle}
\newcommand{\lb}{\rangle}
\newcommand{\ep}{\epsilon}
 \def\we{\wedge}
 \def\lr{\leftrightarrow}
 \def\f {\frac}
 \def\ti{\tilde}
 \def\ap{\alpha}
 \def\pr{\propto}
 \def\mb{\mathbf}
 \def\ddd{\cdot\cdot\cdot}
 \def\no{\nonumber \\}
 \def\la{\langle}
 \def\lb{\rangle}
 \def\ep{\epsilon}

\title{\bf Orthogonal Bases of Invariants \\
in \\
Tensor Models}
\author{Pablo Diaz\thanks{pablo.diazbenito@uleth.ca}, \,\,\, Soo-Jong Rey\thanks{rey.soojong@gmail.com} \\
\\
{\small \emph{$^*$Department of Physics and Astronomy, University of Lethbridge,}}\\
{\small \emph{4401 University Drive, Lethbridge, Alberta, T1K 3M4 Canada}}\\
{\small \emph{$^{\dag}$Fields, Gravity \& Strings, {\rm CTPU}}}\\
{\small \emph{Institute for Basic Science, Seoul 06544 Korea}}\\
{\small \emph{$^{\dag}$School of Physics and Astronomy \& Center for Theoretical Physics}}\\
{\small \emph{Seoul National University, Seoul 06544 Korea}}\\
{\small \emph{$^{\dag}$Department of Basic Sciences}} \\
{\small \emph{University of Science and Technology, Daejeon 34113 Korea}}
}

\maketitle
\begin{abstract}
Representation theory provides an efficient framework to count and classify invariants in tensor models of (gauge) symmetry $G _d = U(N_1) \otimes \cdots \otimes U(N_d)$ . We show that there are two natural ways of counting invariants, one for arbitrary $G_d$ and another valid for large rank of $G_d$. We construct basis of invariant operators based on the counting, and compute correlators of their elements. The basis associated with finite rank of $G_d$ diagonalizes two-point function. It is analogous to the restricted Schur basis used in matrix models. We comment on future directions for investigation.
\\

\end{abstract}
\newpage
\tableofcontents
\section{Introduction}
There are various motivations that make a tensor model an interesting system to study. One motivation comes from a scheme for studying quantum entanglement. From the quantum mechanical point of view, rank $d$ tensor models are associated with the multilinear symmetry group $G_d({\bf N}) = U(N_1)\otimes U(N_2) \otimes \cdots \otimes U(N_d)$ acting on a tensor product Hilbert space $\mathcal{H}=\mathcal{H}_{N_1}\otimes \cdots\otimes \mathcal{H}_{N_d}$. We know that the Hilbert space of a composed physical system is the tensor product of its constituents, this is an essential aspect of entanglement in quantum mechanics \cite{EPR}. So tensor models naturally describe composite systems. Moreover, gauge invariant operators built out of tensors separate the entangled and unentangled states of $\mathcal{H}$, so they can be viewed as a probe of quantum entanglement measurements \cite{CM}.

Another motivation comes from a scheme for studying quantum gravity. Inspired by the success of matrix models in describing two-dimensional quantum gravity \cite{earlierTM}, tensor model was proposed as a framework for describing higher-dimensional random geometry \cite{random1,random2,random3}.
The colored tensor models \cite{color1,color2} and the development of its  $1/N$-expansion \cite{1N1,1N2,1N3} triggered a fast growth of the field of tensor model in recent years. The introduction of color has served to overcome several difficulties that the earlier tensor models had in describing quantum gravity at dimensions greater than two. More recently, the colored tensor model have  been found in direct connection with the $\text{AdS}_2/\text{CFT}_1$ holography, through an alternative formulation of the Sachdev-Ye-Kitaev (SYK) model \cite{SYK1,SYK2,SYK3,SYK4,SYK5,SYK6} in which the necessity of quenched disorder is dispensed \cite{Witten}, see also \cite{Razvan}.

The simplest yet nontrivial tensor model is the matrix model, which was recently studied extensively in the context of AdS/CFT correspondence. In the matrix model, the use of orthogonal bases for two-point functions (first for the BPS-sector \cite{CJR} and then for general bosonic sectors \cite{restricted1,restricted2,restricted3,brauer1,brauer2,Brown:2007xh,Brown:2008ij} and involving gauge field \cite{restricted4} or fermions \cite{restricted5}) was extremely useful for computations in $\mathcal{N}=4$ super Yang-Mills theory within the so-called non-planar regime, which involves heavy operators dual to excited D-branes and solitonic objects in the string theory side \cite{applications1,applications2,applications3,applications4,applications5}. 

The aim of this paper is to set an analogous framework for tensor models. We first count tensor invariants, following the steps of \cite{HW} and \cite{GR}. We then construct bases of invariants which diagonalize the two-point functions and we finally compute exact correlators of the elements of the given basis.
We argue that representation theory provides two natural ways of counting gauge invariant tensor operators. One is valid for arbitrary rank of the symmetry group $G_d$, while the other is only valid at large rank of it. In Section \ref{Counting}, we explore both methods of counting. Guided by them, in Section \ref{basis}, we construct bases of  gauge invariant operators and propose a basis for finite rank of symmetry group $G_d$ that diagonalizes the free two-point function of the tensor model. In section \ref{corre}, we compute the correlators of basis elements. Some directions for future study are discussed in Section \ref{Conclusion}.

\section{Two Methods of Counting Invariants}\label{Counting}
Colored tensors are tensors with no further symmetry assumed. A $d$ covariant color tensor can be written as 
\begin{equation}\label{dtensor}
 \Phi=\Phi_{i_1i_2\dots i_d}~e^{i_1}\otimes e^{i_2}\otimes \cdots \otimes e^{i_d},
\end{equation}
 where $\{e^{i_k}\}$ form a basis of $\mathbb{C}^{N_k}$, so $i_k=1,\dots, N_k$. The objects  $ \Phi_{i_1i_2\dots i_d}$ transform under the action of $G_d =U(N_1)\otimes U(N_2)\otimes \cdots \otimes U(N_d)$ as
\begin{equation}\label{unitaryaction}
 \Phi_{j_1j_2\dots j_d}=\sum_{i_1,\dots, i_d}U(N_1)_{j_1}^{i_1}U(N_2)_{j_2}^{i_2} \cdots U(N_d)_{j_d}^{i_d}  \Phi_{i_1i_2\dots i_d}.
\end{equation}
The complex conjugate is a contravariant tensor that transforms as
\begin{equation}\label{unitaryactionconjugate}
 \overline{\Phi}^{j_1j_2\dots j_d}=\sum_{i_1,\dots, i_d}\overline{U}(N_1)^{j_1}_{i_1}\overline{U}(N_2)^{j_2}_{i_2} \cdots \overline{U}(N_d)^{j_d}_{i_d}  \overline{\Phi}^{i_1i_2\dots i_d}.
\end{equation}

We will be interested in the $n$-fold tensor product $\Phi^{\otimes n}$, built out of $n$ copies of Eq.(\ref{dtensor}). For these objects, we will use indices $i^p_k$ where $p=1,\dots,n$ and $k=1,\dots,d$. So, a basis of $\Phi^{\otimes n}$ can be written as
\begin{equation}
\bigotimes_{p=1}^n\bigotimes_{k=1}^d e^{i^p_k}\qquad \mbox{where} \qquad i^p_k=1,\dots, N_k.
\end{equation}
Note that the group $G_d$ acts diagonally ($n$ times) on $\Phi^{\otimes n}$. Now, as we want the copies to be indistinguishable, we will take the average $\text{Sym}(\Phi)^{\otimes n}$. For fixed $n$, consider operators  of the form
\begin{equation}
\mathcal{O}=\text{Sym}(\Phi)^{\otimes n}\otimes\text{Sym}(\overline{\Phi})^{\otimes n},
\end{equation}
and select the set of these operators which are invariant under the action of $G_d$. They will be referred to as $\mathcal{O}^{G_d-\text{inv}}$.

We first observe that invariants of tensors under the simultaneous unitary action (\ref{unitaryaction}) and (\ref{unitaryactionconjugate}) are obtainable from contracting in all possible ways pairs of covariant and contravariant tensors. In other words, the set
\begin{equation}\label{spanset}
\Big\{\mathcal{O}_{\alpha_1\dots\alpha_d}=\prod_{p=1}^n\Phi_{i_1^pi_2^p\dots i_d^p}\overline{\Phi}^{i_1^{\alpha_1(p)}i_2^{\alpha_2(p)}\dots i_d^{\alpha_d(p)}}|\alpha_1,\dots\alpha_d\in S_n\Big\}
\end{equation}
spans the space of invariants. This is so because the space of $U(N_k)$-invariant linear maps
\begin{equation}\label{map}
\iota: e_i\otimes e^j\to \delta_{i}^j
\end{equation}
is one-dimensional and, as we have $n$ copies of both $\Phi$ and $\overline{\Phi}$, the map (\ref{map}) can apply to any of the permuted slots. Obviously, this holds for each tensor index, resulting in $d$ permutations of $n$ elements for an $n$-fold product of a rank-$d$ tensor, as shown in the set (\ref{spanset}).

Note that, though every invariant can be expressed as a linear combination of the elements of (\ref{spanset}), the set (\ref{spanset}) does not form a basis simply because the elements are not linearly independent. The first problem is to find a way of counting the number of $n$-fold invariants of rank $d$ tensors. Applying arguments from representation theory, we will find two natural ways of counting invariants, one that applies to arbitrary ranks $N_k$ of the constituent unitary groups and the other that holds for large ranks $N_k$, more specifically, for $N_k\geq n$ for all $k$. This problem was independently addressed in \cite{HW} and \cite{GR}. We will study them first and use the labels of these two ways of counting to construct the respective bases of invariants.

\subsection{Finite rank $N_k$}
   
Call $V_n$ and $\overline{V}_n$  the vector spaces spanned by $\text{Sym}(\Phi)^{\otimes n}$ and $\text{Sym}(\overline{\Phi})^{\otimes n}$, respectively.
The action of the group $G_d$ on $\mathcal{O}$ is defined by its simultaneous diagonal action on  both $\Phi^{\otimes n}$ and $\overline{\Phi}^{\otimes n}$. This action will split $V_n$ and $\overline{V}_n$, which are isomorphic each other, into representations of $G_d =U(N_1)\otimes U(N_2)\otimes \cdots \otimes U(N_d)$. Consider the index $k$ out of the $d$ indices of $\Phi$. In the $n$-fold product $\Phi^{\otimes n}$, the space associated with this index is isomorphic to $(\mathbb{C}^{N_k})^{\otimes n}$. Now, as a consequence of Schur-Weyl duality, irreducible representations of $(\mathbb{C}^{N_k})^{\otimes n}$ under the diagonal action of $U(N_k)$ are labeled by Young diagrams with $n$ boxes with at most $N_k$ rows.  Thus, the irreducible representations of $V_n$ (and of $\overline{V}_n$ by the isomorphism) under the action of  $G_d$ are labeled by collections $(\mu_1,\dots,\mu_d)$, where $\mu_k$ are Young diagrams with $n$ boxes, denoted as $|\mu_k|=n$. The number of rows of each diagram does not exceed $N_k$, that is, $l(\mu_k)\leq N_k$.

The problem of classifying $\mathcal{O}^{G_d-\text{inv}}$, the $G_d$-invariants of $V_n\otimes \overline{V}_n$, translates into a representation theory problem since the invariants are in one-to-one correspondence with $G_d$-invariant maps $(V_n, \overline{V}_n)\to \mathbb{C}$, that is,
\begin{equation}\label{numberofinvariants} 
\text{dim}\{\mathcal{O}^{G_d-\text{inv}}\}=\text{dim}\text{ Hom}_{G_d}(V_n,\overline{V}_n),
\end{equation}
and, by Schur's Lemma, there exists one homomorphism (modulo an equivalence) every time we pair up an irreducible representation (irrep) of $V_n$ with an irrep of $\overline{V}_n$.\\
Denote $N=N_1N_2\cdots N_d$. It is clear that one can map $\otimes_{i=1}^d \mathbb{C}^{N_i}\to \mathbb{C}^N$. This is called the Kronecker map, and produces an embedding of the Kronecker product of matrices $\otimes_{i=1}^dU(N_i)$ into $U(N)$. In turn, this maps 
\begin{equation}
V_n\to R_{(n)}^N,
\end{equation}
as $R_{(n)}^N\cong \text{Sym}(\mathbb{C}^{N})^{\otimes n}$ from the Schur-Weyl duality\footnote{The Schur-Weyl duality asserts that $(\mathbb{C}^{N})^{\otimes n}=\oplus_\lambda R^N_\lambda \otimes \Gamma_\lambda$ under the action of $U(N)$ and $S_n$, where $R^N_\lambda$ and $\Gamma_\lambda$ are irreps of $U(N)$ and $S_n$, respectively. The operation ``Sym'' projects the direct sum into the subspace labeled by $\lambda=(n)$. As $\Gamma_{(n)}$ is one-dimensional, it follows that $R_{(n)}^N\cong \text{Sym}(\mathbb{C}^{N})^{\otimes n}$.}. The decomposition of a general irrep $R^N_{\mu}$ of $U(N)$ under the Kronecker map just defined is known. For $|\mu|=n$, one has
\begin{equation}\label{Kroneckerunitary}
R_{\mu}^N=\bigoplus_{\substack{|\mu_1|,\dots,|\mu_d|=n\\
l(\mu_k)\leq N_k}}g_{\mu_1,\dots \mu_d,\mu}R_{\mu_1}^{N_1}\otimes \cdots \otimes R_{\mu_d}^{N_d},
\end{equation}
where $g_{\mu_1,\dots \mu_d,\mu}$ are the Kronecker coefficients. For the case of interest, $\mu=(n)$. Now, $g_{\mu_1,\dots \mu_d,(n)}=g_{\mu_1,\dots \mu_d}$, as can be checked by the general formula
\begin{equation}\label{Kroneckergral}
g_{\mu_1,\dots,\mu_d}=\frac{1}{n!}\sum_{\alpha\in S_n}\chi_{\mu_1}(\alpha)\cdots \chi_{\mu_d}(\alpha),\quad \mu_1,\dots,\mu_d\vdash n,
\end{equation}
since $\chi_{(n)}(\alpha)=1$. 

We thus found the decomposition
\begin{eqnarray}\label{decomposition}
V_n&\cong&\bigoplus_{\substack{|\mu_1|,\dots,|\mu_d|=n\\
l(\mu_k)\leq N_k}}g_{\mu_1,\dots,\mu_d}R^{N_1}_{\mu_1}\otimes\cdots\otimes R^{N_d}_{\mu_d},\nonumber \\
\overline{V}_n&\cong&\bigoplus_{\substack{|\mu_1|,\dots,|\mu_d|=n\\
l(\mu_k)\leq N_k}}g_{\mu_1,\dots,\mu_d}\overline{R}^{N_1}_{\mu_1}\otimes\cdots\otimes \overline{R}^{N_d}_{\mu_d},
\end{eqnarray}
where the representation $\overline{R}_{\mu_k}$ is isomorphic to the irrep $R_{\mu_k}$ in the contravariant basis.  The Kronecker coefficients $g_{\mu_1,\dots,\mu_d}$ are thus the multiplicity of irrep  $(\mu_1,\dots,\mu_d)$ in the decomposition. Equivalently, $g_{\mu_1,\dots,\mu_d}$ is the number of orbits labeled by $(\mu_1,\dots,\mu_d)$ that appear in $V_n$ when acted on by $G_d$. 

We now can apply the decomposition (\ref{decomposition}) into Eq.(\ref{numberofinvariants}) and obtain the formula
\begin{tcolorbox}
\begin{equation}\label{numberofinvariantscomplete} 
\text{dim}\{\mathcal{O}^{G_d-\text{inv}}\}=\text{dim}\text{ Hom}_{G_d}(V_n,\overline{V}_n)=\sum_{\substack{|\mu_1|,\dots,|\mu_d|=n\\
l(\mu_k)\leq N_k}}g^2(\mu_1,\dots,\mu_d).
\end{equation}
\end{tcolorbox}
\noindent This formula agrees with the result found in \cite{HW}. In the table (\ref{table}), we illustrate this  result by enlisting the number of invariants for some values of $n$ and $N_1=N_2=N_3\equiv N$, for the case $d=3$.

\begin{equation}\label{table}
\begin{tabular}{|c|c|c|c|c|c|}
\hline
& $N$=1&$N$=2&$N$=3&$N$=4&$N$=5 \\
\hline
$n$=1&~1&1&1&1&1\\
\hline
$n$=2&~1&4&4&4&4\\
\hline
$n$=3&~1&5&11&11&11\\
\hline
$n$=4&~1&12&31&43&43\\
\hline
$n$=5&~1&15&92&143&161\\
\hline
\end{tabular}
\end{equation}

\subsection{Large rank $N_k$}
If  $N_k$ were large enough, viz. $N_k\geq n$ for all $k$, there exists an alternative way of counting invariants, based on the observation that all invariants is expressible as linear combinations of elements in the set (\ref{spanset}), subject to equivalence of a double diagonal action of $S_n$. This is so because the initial ordering of the $n$ slots in $\Phi^{\otimes n}$ and in $\overline{\Phi}^{\otimes n}$ is irrelevant after symmetrizing. So, the number of invariants coincides with the size of double coset
\begin{equation}\label{dc}
\text{Diag}(S_n)\backslash S_n^{\times d} /\text{Diag}(S_n).
\end{equation}
The size of double coset (\ref{dc}) can be calculated using Burnside's Lemma \cite{GR,HW}. It results in the simple formula
\begin{tcolorbox}
\begin{equation}\label{numberofinvariants2}
\text{dim}\{\mathcal{O}^{G_d-\text{Inv}}\}=|\text{Diag}(S_n)\backslash S_n^{\times d} /\text{Diag}(S_n)|=\sum_{\lambda\vdash n}z_{\lambda}^{d-2},
\end{equation}
\end{tcolorbox}
\noindent
where $z_\lambda$ is combinatorial number that depends on the partition $\lambda$ of $n$ as follows. If we write the partition $\lambda=(\lambda_1,\dots,\lambda_n)$ such that $n=\sum_i i\lambda_i$, then
\begin{equation}
z_\lambda=\prod_{i=1}^n i^{\lambda_i}(\lambda_i!).
\end{equation}
The formula (\ref{numberofinvariants2}) is much simpler than the formula (\ref{numberofinvariantscomplete}). Actually, computing Eq.(\ref{numberofinvariantscomplete}) rapidly becomes out of reach as $n$ grows, since there is no combinatorial method available to date for computing Kronecker coefficients. 

One can readily check that both formula agree each other. Evaluating Eq. (\ref{numberofinvariants2}) for $d=3$ and $n=1,2,3,4,5$, we get $1,4,11,43,161$. We see that they match with the last column of Table (\ref{table}). The general proof that both formulas coincide for large $N_k$ can be found in Proposition 5 of \cite{HW}. The idea is that, besides (\ref{Kroneckerunitary}),  Kronecker coefficients also appear in the Kronecker product of irreps of $S_n$ as\footnote{The proof that $g_{\mu_1\dots\mu_d\mu}$ in (\ref{Kroneckerunitary}) are the same numbers as in (\ref{Kroneckersymmetric}) relies on Schur-Weyl duality. }
\begin{equation}\label{Kroneckersymmetric}
\Gamma_{\mu_1}\otimes \cdots \otimes\Gamma_{\mu_d}=\bigoplus_{\mu}g_{\mu_1\dots\mu_d\mu}\Gamma_{\mu}.
\end{equation}
using this fact, the size of the double coset (\ref{dc}) can be proven to be 
\begin{equation}\label{kroneckerdc}
|\text{Diag}(S_n)\backslash S_n^{\times d} /\text{Diag}(S_n)|=\sum_{|\mu_1|,\dots,|\mu_d|=n}g_{\mu_1\dots\mu_d}^2.
\end{equation}
The difference between (\ref{kroneckerdc}) and (\ref{numberofinvariantscomplete}) is that in the counting (\ref{kroneckerdc}) there is no restriction in the number of columns of the irreps. This happens because (\ref{kroneckerdc}) is derived from (\ref{Kroneckersymmetric}). As a consequence, the formula derived from the double coset counts the number of invariants only for large $N_k$, otherwise it overestimates it.

\section{Bases of Invariant Operators}\label{basis}
We next move to construct explicit bases of the invariants. The counting methods we developed in the previous section will serve as a guidance  for the construction. We will see that, associated with the two ``natural'' counting methods, it is possible to construct two types of bases.

  Let us start with the case of finite $N_k$. The relevant formula is Eq.(\ref{numberofinvariantscomplete}). From this formula we learn two things:
\begin{itemize}
\item[i)] The first equality of Eq.(\ref{numberofinvariantscomplete}) tells us that there exists one invariant operator every time we couple an irrep of $V_n$ with its dual in $\overline{V}_n$. If we associate each irrep of $V_n$ with a vector, then invariants are in one-to-one correspondence with vectors for the subspace of $V_n$ where there is no multiplicity. In the subspaces for which a certain irrep occurs more than once, invariants are in one-to-one correspondence with endomorphisms. For example, if a certain irrep occurs twice, there are four ways of pairing: $\{(v_1,\overline{v}_1),(v_1,\overline{v}_2),(v_2,\overline{v}_1),(v_2,\overline{v}_2)\}$.
\item[ii)] The second equality of Eq.(\ref{numberofinvariantscomplete}) tells us precise information about the decomposition of $V_n$ and the suitable labels to describe it. As can be read from of Eq.(\ref{numberofinvariantscomplete}), the set of labels that exhausts the counting is $\{\mu_1,\dots,\mu_d,ij\}$, where $\mu_k\vdash n$ with $l(\mu_k)\leq N_k$, and  $i,j=1,\dots,g_{\mu_1\dots\mu_d}$. 
\end{itemize}
As a basis of invariant operators for finite $N_k$, we propose
\begin{equation}\label{operatorsfiniteN}
\mathcal{O}_{\mu_1\dots\mu_d,ij}=\text{Tr}\big(V_n\mathcal{P}_{\mu_1\dots\mu_d,ij}\overline{V}_n\big),
\end{equation}
where $\text{Tr}$ is an instruction to contract all the tensor indices of the elements of $V_n$ with those of $\overline{V}_n$ such that the result is an invariant. Here, $\mathcal{P}_{\mu_1\dots\mu_d,ij}$ is the projector that acts on the vector space $V_n$ and projects onto the subspace labeled by $\mu_1\dots\mu_d$ (which has multiplicity $g_{\mu_1\dots\mu_d}$). As a basis of endomorphisms, we choose intertwiners labeled by $i,j$. So\footnote{Note the similarity of the basis so constructed with the restricted Schur basis built on matrix models \cite{restricted1,restricted2,restricted3}.},
\begin{eqnarray}
\mathcal{P}_{\mu_1\dots\mu_d,ij}\mathcal{P}_{\mu'_1\dots\mu'_d,i'j'}&=&\delta_{\mu_1\mu'_1}\cdots\delta_{\mu_d\mu'_d}\delta_{ji'}
\mathcal{P}_{\mu_1\dots\mu_d,ij'}\nonumber \\
\sum_{\mu_1\dots\mu_d}\sum_{i=1}^{g_{\mu_1\dots\mu_d}}\mathcal{P}_{\mu_1\dots\mu_d,ii}&=&\mathbb{1}.
\label{above}
\end{eqnarray}
In view of the decomposition Eq.(\ref{decomposition}), the operators (\ref{operatorsfiniteN}) can be equivalently written as
\begin{tcolorbox}
\vskip-0.5cm
\begin{eqnarray}\label{operatorsfiniteNdouble}
&& \mathcal{O}_{\mu_1\dots\mu_d,ij}=\text{Tr}\big(\Phi_{\mu_1\dots\mu_d,i}\overline{\Phi}_{\mu_1\dots\mu_d,j}\big),\nonumber \\
&& \overline{\mathcal{O}}_{\mu_1\dots\mu_d,ij}=\text{Tr}\big(\Phi_{\mu_1\dots\mu_d,j}\overline{\Phi}_{\mu_1\dots\mu_d,i}\big),
\end{eqnarray}
\end{tcolorbox}
\noindent where we have referred to $\Phi_{\mu_1\dots\mu_d,i}$ and $\overline{\Phi}_{\mu_1\dots\mu_d,j}$ for the subspaces of $V_n$ and $\overline{V}_n$ corresponding to copy $i$ and copy $j$, respectively, of the irrep labeled by   $(\mu_1,\dots,\mu_d)$.

Projectors on the labels $\mu_1,\dots,\mu_d$ can be constructed as follows. Start from the standard projectors,
\begin{equation}\label{Pmu}
P^{\mu}=\frac{d_{\mu}}{n!}\sum_{\sigma\in S_n}\chi_{\mu}(\sigma)\sigma,
\end{equation}
which projects the tensor product $(\mathbb{C}^N)^{\otimes n}$ onto the subspace $R^N_{\mu}\otimes \Gamma_{\mu}$ of the Schur-Weyl decomposition. Applying the standard projector (\ref{Pmu}) on each index of $\Phi^{\otimes n}$, we then define the projectors
\begin{equation}\label{projectorsisotypical}
\mathcal{P}_{\mu_1\dots\mu_d}\equiv\frac{d_{\mu_1}\cdots d_{\mu_d}}{n!^d}\sum_{\sigma_1,\dots,\sigma_d\in S_n} \chi_{\mu_1}(\sigma_1)\cdots \chi_{\mu_d}(\sigma_d)\sigma_1\cdots \sigma_d,
\end{equation}
where each permutation acts on a different index of $\Phi$. These projectors are then related to the projector $\mathcal{P}_{\mu_1\dots\mu_d,ij}$ in Eqs.(\ref{operatorsfiniteN}, \ref{above}) as
\begin{equation}
 \mathcal{P}_{\mu_1\dots\mu_d}=\sum_{i=1}^{g_{\mu_1\dots\mu_d}}\mathcal{P}_{\mu_1\dots\mu_d,ii},
\end{equation}
that is, $ \mathcal{P}_{\mu_1\dots\mu_d}$ projects on the isotypical component. Associated with projectors (\ref{projectorsisotypical}), we construct the invariant operators
\begin{equation}\label{operatorsprojection}
\mathcal{O}_{\mu_1\dots\mu_d}=\frac{d_{\mu_1}\cdots d_{\mu_d}}{n!^d}\sum_{\alpha_1,\dots,\alpha_d\in S_n} \chi_{\mu_1}(\alpha_1)\cdots \chi_{\mu_d}(\alpha_d)\mathcal{O}_{\alpha_1\dots\alpha_d},
\end{equation}
where $\mathcal{O}_{\alpha_1\dots\alpha_d}$'s are as in Eq.(\ref{spanset}).
In general, operators $\mathcal{O}_{\mu_1\dots\mu_d}$ do not form a basis, except for special cases like $d=3$ and $n=1,2,3,4$, where there are no multiplicities and so they coincide with $\mathcal{O}_{\mu_1\dots\mu_d,ij}$. However, we have an explicit construction of them and, as we will shown below, we find that they form an orthogonal set of the two-point function. 
An explicit construction of $\mathcal{O}_{\mu_1\dots\mu_d,ij}$ in terms of permutations must exist since, as discussed before, the set (\ref{spanset}) spans the space of invariants operators. We leave it for a future work. 

Alternative bases of invariant operators can be constructed in the case that $n\leq N_k$ for all $k$. In the spirit of the double coset counting, two invariant operators $\mathcal{O}_{\alpha_1\dots\alpha_d}$ and $\mathcal{O}_{\beta_1\dots\beta_d}$ are linearly independent if and only if it does not exist $\tau,\sigma\in S_n$ such that $\tau\alpha_i\sigma=\beta_i$ for all $i$ \footnote{Note that this condition does not guarantee linear independence if $n>N_k$ for any $k$.}. Now, for every monomial $\mathcal{O}_{\alpha_1\dots\alpha_d}$, we can choose a representative multiplying all the permutations by $\alpha_d^{-1}$. So, after reordering, we are left with a collection of operators
\begin{equation}
\{\mathcal{O}_{\beta_1\dots\beta_{d-1}1}|\beta_1,\dots,\beta_{d-1}\in S_n\}.
\end{equation}
These operators still have the equivalence
\begin{equation}
\mathcal{O}_{\beta_1\dots\beta_{d-1}1}\quad \sim \quad \mathcal{O}_{\tau\beta_1\tau^{-1}\dots\tau\beta_{d-1}\tau^{-1}1},
\end{equation}
otherwise, they are linearly independent. Now we choose representatives of the orbits of $(\beta_1,\dots\beta_{d-1})$ generated by simultaneous conjugation. Each representative will be a collection $(\sigma_1,\dots,\sigma_{d-1})$. Then, the set 
\begin{tcolorbox}
\begin{equation}\label{basislargeN}
\{\mathcal{O}_{\sigma_1\dots\sigma_{d-1}1}|(\sigma_1,\dots,\sigma_{d-1}) \text{  representative}\}
\end{equation}
\end{tcolorbox}
\noindent forms a basis.

On general grounds, we do not expect that the basis (\ref{basislargeN}) is orthogonal under the two-point function. So, it will only have a limited utility for computations. A clear advantage of providing an orthogonal basis with easy expressions for the correlators is that it serves to compute correlators of generic observables, as they can always decompose into linear combinations of the elements of the  basis. Thus, it will be desirable to build an orthogonal basis for the large $N_k$ case. Here, we sketch how to do so, leaving detailed study to our forthcoming work \cite{DR}. The idea is to focus on the counting (\ref{numberofinvariants2}). We see that the number of invariants is counted as a sum of partitions $\lambda$ and the value of $z_{\lambda}$. The key observation is that $z_{\lambda}$ counts the number of permutations that commute with a given permutation $x$, which has cycle structure $\lambda$. In other words, given $x$ with $[x]=\lambda$, $z_{\lambda}$  is the number of solutions of the equation 
\begin{equation}\label{zlambdasolutions}
\sigma x\sigma^{-1}=x, \quad \sigma \in S_n.
\end{equation}
For instance, if $x$ is the identity, then there are $n!$ solutions since every permutation would solve the equation.
Solutions of Eq.(\ref{zlambdasolutions}) form a subgroup $H_x\subset S_n$. The structure of $H_x$ can be read off from the diagram $\lambda$ in this way: If we write the partition $\lambda=(\lambda_1,\dots,\lambda_n)$ so that $n=\sum_i i\lambda_i$, then
\begin{equation}
H_x=\times_{i=1}^n~S_{\lambda_i}\wr C_i, \quad [x]=\lambda,
\end{equation}
where $S_{\lambda_i}\wr C_i$ is the wreath product of $S_{\lambda_i}$ with the cyclic group of size $i$. The idea is to use the subgroup $H_x$ to form an orthogonal basis. For more details and explicit constructions, see \cite{DR}.

\section{Correlators}\label{corre}
Consider a free tensor model, defined by the partition function,
\begin{equation}
Z=\int d\Phi d\overline{\Phi}e^{-\Phi \cdot \overline{\Phi}}.
\end{equation}
Here, in the probability distribution function, the quadratic term $\Phi\overline{\Phi}$ is chosen to be the simplest 
\begin{equation}
\Phi \cdot \overline{\Phi}=\Phi_{i_1\dots i_d}\overline{\Phi}^{i_1\dots i_d},
\end{equation}
with repeated indices contracted. The two-point correlator of this model reads
\begin{equation}
\langle \Phi_{i_1\dots i_d}\overline{\Phi}^{j_1\dots j_d}\rangle={1 \over Z} \int d\Phi d\overline{\Phi}~\Phi_{i_1\dots i_d}\overline{\Phi}^{j_1\dots j_d}e^{-\Phi\overline{\Phi}}=\delta_{i_1}^{j_1}\cdots \delta_{i_d}^{j_d}.
\end{equation}
If we have $n$ copies of $\Phi$ and $\overline{\Phi}$, then we get a sum over Wick contractions
\begin{equation}
\langle \prod_{p=1}^n\Phi_{i^p_1\dots i^p_d}\prod_{q=1}^n\overline{\Phi}^{j^q_1\dots j^q_d}\rangle=\sum_{\sigma\in S_n}\prod_{p=1}^n \delta_{i_1^p}^{j_1^{\sigma(p)}}\cdots \delta_{i_d^p}^{j_d^{\sigma(p)}}.
\end{equation}
The invariant operators we are considering here have the schematic structure $\mathcal{O}=\Phi^{\otimes n}\otimes \overline{\Phi}^{\otimes n}$. When computing correlators of the form $\langle\mathcal{O}\overline{\mathcal{O}}'\rangle$ we will consider each operator normal ordered, so that we will only allow contractions between $\Phi$'s of  $\mathcal{O}$ and $\overline{\Phi}$'s of $\overline{\mathcal{O}}'$ and between $\overline{\Phi}$'s of $\mathcal{O}$  and $\Phi$'s of $\overline{\mathcal{O}}'$. For this reason, the sum in the correlator $\langle\mathcal{O}\overline{\mathcal{O}}'\rangle$ will be the sum over Wick contractions determined by the two permutations $\sigma,\tau\in S_n$.  \\
For invariant operators of the form (\ref{spanset}), we have
\begin{equation}\label{correlatorsalphas}
\langle\mathcal{O}_{\alpha_1\dots \alpha_d}\overline{\mathcal{O}}_{\beta_1\dots\beta_d}\rangle=\sum_{\sigma,\tau\in S_n}N_1^{C(\sigma\alpha_1\tau\beta_1^{-1})}N_2^{C(\sigma\alpha_2\tau\beta_2^{-1})}\cdots N_d^{C(\sigma\alpha_d\tau\beta_d^{-1})},
\end{equation}
where $C(\sigma)$ is the number of disjoint cycles of permutation $\sigma$. We will use Eq.(\ref{correlatorsalphas}) to compute
the correlators of the bases we proposed in the previous section. 
For explicit  computations, we will need the identity
\begin{equation} \label{cycleexpansion}
N_k^{C(\tau)}=\frac{1}{n!}\sum_{\lambda\vdash n}d_\lambda\chi_\lambda(\tau)f_{\lambda}(N_k),
\end{equation}
which should be read as an explicit expansion of the function $N_k^{C(\tau)}$ (which is a class function since it depends only on the cycle structure of $\tau$)  in terms of characters of the symmetric group which form a basis of class functions \footnote{The formula (\ref{cycleexpansion}) can be derived from the relation between characters of the symmetric group and Schur functions \cite{M}.}. The combinatorial function $f_{\lambda}(N_k)$ is readily constructed from the corresponding Young diagram $\lambda$ as
\begin{equation}
f_{\lambda}(N_k)=\prod_{i,j}(N_k-i+j),
\end{equation}
where $i,j$ are coordinates of the Young diagram $\lambda$ starting from the top left. So, $i$ is the row number and $j$ is the column number. Using Eq.(\ref{cycleexpansion}), we may write the correlators in terms of the characters of the symmetric group and functions $f_{\lambda}(N_k)$ as\footnote{The recent work \cite{MM} also derived an equivalent expression for the correlators.} 
\begin{equation}\label{correlatorsintermsofcharacters}
\langle\mathcal{O}_{\alpha_1\dots \alpha_d}\overline{\mathcal{O}}_{\beta_1\dots\beta_d}\rangle=\frac{1}{n!^d}\sum_{\substack{\sigma,\tau\in S_n\\
\mu_1,\dots,\mu_d\vdash n}}\prod_{k=1}^{d}d_{\mu_k}\chi_{\mu_k}(\sigma\alpha_k\tau\beta_k^{-1})f_{\mu_i}(N_k).
\end{equation}
Now, let us first consider the bases we have proposed at large $N_k$. We will have
\begin{equation}\label{correlatorslargeN}
\langle\mathcal{O}_{\sigma_1\dots \sigma_{d-1}}\overline{\mathcal{O}}_{\overline{\sigma}_1\dots\overline{\sigma}_{d-1}}\rangle=\sum_{\sigma,\tau\in S_n}N_1^{C(\sigma\sigma_1\tau\overline{\sigma}_1^{-1})}\cdots N_{d-1}^{C(\sigma\sigma_{d-1}\tau\sigma_{d-1}^{-1})}N_d^{C(\sigma\tau)},
\end{equation}
where $(\sigma_1,\dots\sigma_{d-1})$ and $(\overline{\sigma}_1,\dots\overline{\sigma}_{d-1})$ are representatives of the orbits produced by simultaneous conjugation of the $d-1$ permutations. As anticipated in the previous section, the elements of this basis are not orthogonal under the free two-point function. Since Eq.(\ref{correlatorslargeN}) admits little simplification, there is not much useful information in these correlators. \\


More interesting are the correlators of operators defined in Eq.(\ref{operatorsprojection}). For those operators, we have
\begin{equation}\label{correlatorsstart}
\langle\mathcal{O}_{\mu_1\dots \mu_{d}}\overline{\mathcal{O}}_{\nu_1\dots\nu_{d}}\rangle=\frac{1}{n!^{2d}}\sum_{\substack{\alpha_1,\dots,\alpha_d\in S_n\\
\beta_1,\dots,\beta_d\in S_n}} \prod_{k=1}^d d_{\mu_k}d_{\nu_k}\chi_{\mu_k}(\alpha_k)\chi_{\nu_k}(\beta_k)\langle\mathcal{O}_{\alpha_1\dots\alpha_d}\overline{\mathcal{O}}_{\beta_1\dots\beta_d}\rangle \, . 
\end{equation}
Let us substitute Eq.(\ref{correlatorsintermsofcharacters}) into Eq.(\ref{correlatorsstart}). Using the orthogonality relation for characters
\begin{equation}
\frac{1}{n!}\sum_{\sigma\in S_n}\chi_{\mu_k}(\sigma)\chi_{\nu_k}(\sigma^{-1}\tau)=\delta_{\mu_k\nu_k}\frac{1}{d_{\mu_k}}\chi_{\mu_k}(\tau)
\end{equation}
 for every $k=1,\dots,d$ in Eq.(\ref{correlatorsstart}), we get 
\begin{eqnarray}\label{correlatorsresult}
\langle\mathcal{O}_{\mu_1\dots \mu_{d}}\overline{\mathcal{O}}_{\nu_1\dots\nu_{d}}\rangle&=&\frac{1}{n!^d}\prod_{k=1}^d\delta_{\mu_k\nu_k}d_{\mu_k}f_{\mu_k}(N_k)\sum_{\sigma\tau\in S_n}\chi_{\mu_k}(\sigma\tau)\nonumber \\
&=&g_{\mu_1\dots\mu_d}\frac{1}{n!^{d-2}}\prod_{k=1}^d\delta_{\mu_k\nu_k}d_{\mu_k}f_{\mu_k}(N_k)\nonumber \\
&=&(n!)^2g_{\mu_1\dots\mu_d}\prod_{k=1}^d\delta_{\mu_k\nu_k}\text{Dim}_{N_k}(\mu_k),
\end{eqnarray}
where $\text{Dim}_N(\mu)$ is the dimension of the irrep $\mu$ of $U(N)$. In these steps, we used Eq.(\ref{Kroneckergral}) and the fact that
\begin{equation}
\text{Dim}_N(\mu)=\frac{d_{\mu}f_{\mu}(N)}{n!}.
\end{equation}
The two-point correlators of the model seems to be perfectly adapted to the classification of the invariants in terms of irreps of $V_n$, in the sense that these invariants are orthogonal under the correlators. These has been proven in Eq.(\ref{correlatorsresult}) at least for the subspaces labeled by $(\mu_1,\dots,\mu_d)$. It still needs to be proven that the basis operators $\mathcal{O}_{\mu_1\dots\mu_d,ij}$ are also orthogonal on the labels $i,j$. Now, since $\mathcal{O}_{\mu_1\dots\mu_d}=\sum_i\mathcal{O}_{\mu_1\dots\mu_d,ii}$, the result Eq.(\ref{correlatorsresult}) suggests that 
\begin{equation}\label{orthogonalelements}
\langle\mathcal{O}_{\mu_1\dots \mu_{d},ij}\overline{\mathcal{O}}_{\nu_1\dots\nu_{d},kl}\rangle=n!^2\delta_{ik}\delta_{jl}\prod_{k=1}^d\delta_{\mu_k\nu_k}\text{Dim}_{N_k}(\mu_k).
\end{equation}
A formal proof of Eq.(\ref{orthogonalelements}) will be relegated in our forthcoming companion work \cite{DR}. Here, we content ourselves with a brief explanation of the idea  why Eq.(\ref{orthogonalelements}) is expected to hold.  We have seen that, because of normal ordering, when we compute correlators $\langle\mathcal{O}\overline{\mathcal{O}}\rangle$, the Wick contractions work independently between the covariant part of $\mathcal{O}$ and the contravariant part of  $\overline{\mathcal{O}}$ and between the contravariant part of $\mathcal{O}$ and the covariant part of $\overline{\mathcal{O}}$. Writing $\mathcal{O}$ and $\overline{\mathcal{O}}$ as in Eq.(\ref{operatorsfiniteNdouble}), we have
\begin{equation}
\langle\mathcal{O}_{\mu_1\dots \mu_{d},ij}\overline{\mathcal{O}}_{\nu_1\dots\nu_{d},kl}\rangle=\langle \text{Tr}\big(\Phi_{\mu_1\dots\mu_d,i}\overline{\Phi}_{\mu_1\dots\mu_d,j}\big)\text{Tr}
\big(\Phi_{\nu_1\dots\nu_d,l}\overline{\Phi}_{\nu_1\dots\nu_d,k}\big)\rangle.
\end{equation}
Orthogonality on the labels $\mu_1,\dots,\mu_d$ follows immediately since the two-point correlator is a $G$-invariant function and the only possible homomorphism between different irreps is null.  Now, the independence of the Wick contractions due to normal ordering and Eq.(\ref{correlatorsresult}) tells us that Eq.(\ref{orthogonalelements}) will hold if 
\begin{equation}\label{orthiso}
\langle \Phi_{\mu_1\dots\mu_d,i}\overline{\Phi}_{\mu_1\dots\mu_d,j}\rangle\quad \sim \quad \delta_{ij},
\end{equation}
that is, if the two-point correlator is also orthogonal for different copies of the isotypical space. So, proving Eq.(\ref{orthiso}) would automatically prove  Eq.(\ref{orthogonalelements}). The proof of Eq.(\ref{orthiso}) will reflect the role of Wick contractions as special $G$-invariant functions. Notice that, in an analogous setup for matrix models (i.e. the restricted Schur basis), the two-point correlator also diagonalizes the operators associated with different components of the isotypical space \cite{restricted1,restricted2,restricted3}.

\section{Summary and future work}\label{Conclusion}
In this work, we used arguments from representation theory to count tensor invariants and to construct bases of them based on the countings. We found two different bases, one valid for arbitrary values of the ranks of symmetry group and a second that counts the number of invariants for large ranks. We computed the correlators of the elements in both bases. The basis associated with the counting at finite rank is analogous to the restricted Schur basis used in matrix models, and it is orthogonal under the two-point correlators of the theory.

Regarding the two countings and the bases, there are two possible extensions of this work. First, it would be interesting to construct an orthogonal basis for large rank of the symmetry group, based on the counting Eq. (\ref{numberofinvariants2}) and perhaps using the arguments given below Eq. (\ref{zlambdasolutions}). Then, we should be able to compare both orthogonal bases, for finite and large ranks, and compute their correlators. Second, it would be useful to establish a rigorous proof of Eq.(\ref{orthogonalelements}) and, if possible, an explicit construction in terms of permutations of the invariants Eq.(\ref{operatorsfiniteN}). All these progresses will be relegated to our forthcoming work \cite{DR}. 

The tensor model we study here is bosonic. If we consider a fermionic tensor model, then we would make contact with the SYK alternative model proposed in \cite{Witten}. To build a fermionic  basis for finite rank, we would start with Eq.(\ref{operatorsfiniteN}) and proceed in an analogous way as was done in  \cite{restricted5} in the context of matrix models. Then, we would be able to perform exact computations for heavy states in the model and compare them with their $\text{AdS}_2$ bulk counterparts \footnote{We would like to thank Robert de Mello Koch for proposing this idea.}.

\vspace{0.2cm}

\section*{\bf Acknowledgment}

\noindent
We thank D. Bak, S. Das, R. de Mello Koch, M. Walton and F. Sugino for useful discussions and valuable feedbacks.  SJR acknowledges participants of the 4th IBS Brainstorm Meeting at Yang-Pyeong, Korea and Erwin Schr\"odinger Institute for Mathematical Physics at Vienna, Austria for many helpful conversations. The work of PD was supported in part by the Natural Sciences and Engineering Research Council of Canada and the University of Lethbridge.

\end{document}